\documentclass[aps,pre, preprint, onecolumn]{revtex4}
\usepackage{amssymb}
\usepackage{eurosym}
\usepackage{epsfig}

\begin{document}

\normalsize

\title{Two stages of magnetic filament formation in the solar  convective zone}
\author{E.\,A. Kuznetsov $^{1,2,3}$, E.\,A. Mikhailov $^{1,4}$, V.\,D.\,Khvoshchinskaia $^4$}
\affiliation{$^{1}$ \textit{P.N. Lebedev Physical Institute of RAS, Moscow, Russia}\\
$^{2}$ \textit{L.D. Landau Theoretical Physics Institute of RAS, Chernogolovka, Russia}\\
$^{3}$ \textit{Space Research Institute of RAS, Moscow, Russia}\\
$^{(4)}$ \textit{M.V. Lomonosov Moscow State University, Moscow, Russia}}

\begin{abstract}
This paper presents a brief overview of studies of magnetic filament formation in the solar convection zone. Two stages of magnetic filament formation and development are distinguished. The first stage can be described within the kinematic approximation for the MHD equation, since the average kinetic energy of convective motion exceeds the average magnetic field energy. Moreover, magnetic Reynolds number is on the order of . Therefore, at the initial stage of magnetic filament formation, the magnetic field can be considered frozen-in. It turns out that at this stage, magnetic filaments begin to form at the boundaries of convective cells in hyperbolic regions of the flow. These regions act as attractors for the magnetic field. At the free boundary of the convective zone near the interfaces between convective cells, the magnetic field has a predominantly normal component relative to the boundary. As the magnetic field increases, the field's frozen-in nature in the filament is disrupted, and saturation occurs due to finite conductivity. It is important to note that the magnetic field growth and saturation in the filament at this stage can be determined by analyzing the behavior of the magnetic field at the free boundary, which, according to observations, can be considered flat. Moreover, the formation of magnetic filaments is independent of the internal structure of the convective cells and is determined by the behavior of the convective flow along the free surface. In the next stage, when the magnetic field in the filament is sufficiently strong, the magnetic field pressure gradient begins to impede the convective flow, leading to a shift of the hyperbolic regions toward the convective flow along the free surface. As a result, the transverse size of the filament increases and stops when the kinetic energy density and the magnetic field energy density become comparable.

\end{abstract}

\maketitle

\section{Introduction}

The drift of charged particles in a high-frequency (HF) electromagnetic field, as is well known from the classic works of A.V. Gaponov and M.A. Miller \cite{GaponovMiller1, GaponovMiller2}, arises due to the HF pressure gradient -- the Gaponov-Miller force. This fact, discovered in 1958, has become a cornerstone concept in the physics of HF waves in plasma. For example, for Langmuir waves, HF pressure pushes the plasma out of the region of electric field localization, forming caverns of reduced plasma density in which the local plasma frequency is lower than the average plasma frequency. This instability, discovered by A.A. Vedenov and L.I. Rudakov \cite{VedenovRudakov}, is known as modulation instability. This results in bound localized states of the Langmuir wave field, for which the adiabatic invariant -- the number of HF waves -- is conserved. However, this process -- the decrease in density due to high-frequency pressure -- can be compensated for by the dispersion of Langmuir waves. Complete compensation is achieved for one-dimensional Langmuir solitons, which are stable with respect to one-dimensional perturbations. In a three-dimensional situation, the formation of an increasingly deeper potential well due to the decrease in density in the cavity does not stop, and the system enters the Langmuir wave collapse regime predicted by V.E. Zakharov \cite{Zakharov} in 1972 (see also \cite{ZakharovKuznetsov}).

This is a striking example of the formation of nonlinear coherent structures in the form of solitons or collapses as a result of the interaction of electromagnetic fields with plasma. In this review, we will examine another example, opposite to Langmuir waves, where the hydrodynamic plasma flow caused by solar convection leads to the generation of localized objects in the form of magnetic filaments, in which the magnetic field significantly exceeds the average solar magnetic field. The emergence of magnetic filaments, as will be demonstrated in this review, is due to the structure of the convective zone of the Sun, which consists of a set of convective cells, in the center of which the liquid floats up, and at the boundaries of the cells it sinks down.

\subsection{Features of frozen fields}

An important role in the formation of magnetic filaments is played by the frozenness  of the magnetic field in magnetohydrodynamics (MHD) at a vanishingly small magnetic viscosity. It should be noted that the frozenness property of vorticity in an ideal fluid was understood already in the 19th century by the classics of hydrodynamics: Cauchy, Kelvin, Helmholtz, and others (see, for example, \cite{Lamb, Batchelor, Salmon, ZakharovKuznetsov1997}). Notice that for a long time the outstanding achievement of Cauchy (see, for example, the book \cite{Lamb}), who found Lagrangian invariants describing the conservation of local vorticity, remained unnoticed. A simple consequence of these invariants is Kelvin's theorem on the conservation of vorticity (obtained almost 40 years after the discovery of Cauchy invariants).  We owe the revival of Cauchy's name in this fundamental question to E.I. Yakubovich and his co-authors (see the articles \cite{Abrashkin, Yakubovich} and references therein).

In ideal MHD, the frozenness of magnetic field was discovered by Alfven in 1942 \cite{Alfven}.
In this review, we will show that the formation of magnetic filaments in the solar convection zone is largely due to the frozen-in magnetic field. Magnetic filaments on the Sun are localized structures with an elevated magnetic field, significantly exceeding the average magnetic field, which is on the order of 10~G.

	In terms of hydrodynamics, the classic works on developed turbulence by Kolmogorov and Obukhov ~\cite{Kolmogorov41,Obukhov41} suggest that, as a result of energy transfer from the large-scale region to the viscous region, the magnitude of fluctuations of vorticity, as a frozen field in the inertial interval, increases at small scales as $\ell^{-2/3}$, which corresponds to the emergence of a singularity. If we assume that the formation time of a singularity is finite, determined by the rate of energy dissipation and the energy-containing scale, then we can say that the emergence of such a singularity can be considered an example of collapse.
 This issue, however, remains controversial. In particular, therefore, one of the main questions in the theory of developed hydrodynamic turbulence is the question of collapse: does the formation of singularities occur in a finite time or not? For example, in the two-dimensional case, the vorticity cannot go to infinity in a finite time. The possibility of collapse in three dimensions is still a matter of debate. Early numerical studies initially suggested its occurrence, but a more thorough examination of the issue cast doubt on these results. There are studies that have numerically demonstrated the possibility of singularity formation on a solid wall for ideal fluids in three dimensions. Collapse does not exhaust all the possibilities for singularity formation. They can arise through exponential growth, i.e., formally, over an infinite time. Such phenomena are observed in numerical simulations when there is exponential growth of the main flow parameters, namely, velocity gradients and vorticity. Such objects include pancake-type vortex structures arising from the excitation of turbulence, which contract exponentially with a simultaneous increase in vorticity ~\cite{Kuznetsov07,Agafontsev22}. The compression of such structures is associated with the frozenness of vortex lines, despite the fact that the vorticity divergence is identically equal to zero. It turns out that a Kolmogorov-type relationship arises between the pancake thickness $\ell$ and the maximum vorticity $\omega_{max}$: $\omega_{max}\propto \ell^{-2/3}$ \cite{Agafontsev15,Agafontsev16}. The same behavior is observed for two-dimensional flows of an ideal fluid ~\cite{Kuznetsov98,Kuznetsov02}. The $2/3$ law is one consequence of frozenness for vorticity in three-dimensional flows and divorticity for two-dimensional hydrodynamics ~\cite{Kuznetsov19}.

In magnetohydrodynamics, similar processes also play an important role both in experiments with sufficiently cold plasmas and in numerous astrophysical phenomena. It is well known that convection plays a significant role in the evolution of magnetic fields in accretion disks surrounding compact objects ~\cite{Shakura73, Rudiger95, Ghasemnezhad17, Boneva21}. Convection is also significant in the development of magnetorotational instability in astrophysics ~\cite{Hawley01}. In this paper, we will study in detail the processes of the formation and evolution of magnetic filaments in the solar convection zone.

As was first shown in the classical works of E.N. Parker \cite{Parker63} , a field enhancement at the boundaries of convective cells in the Sun's convective zone is possible. He demonstrated that the magnetic field in the case of a two-dimensional flow in the form of a periodic lattice of rolls grows exponentially during the generation of magnetic filaments. These ideas were subsequently developed extensively (see \cite{Galloway} - \cite{Ryutova}). Many of these results are presented in the book by Stix \cite{Stix02}, one of the classics in this field. In this regard, we would like to highlight the work of Getling \cite{Getling10} on the generation of magnetic filaments by a convective system of hexagonal cells.

In one of our previous studies, we numerically simulated the corresponding process within the kinematic approximation ~\cite{Kuznetsov20}, when the magnetic field energy is relatively small compared to the energy of turbulent motions. In this case, at the interface between convective cells there are formed  filaments of magnetic field when their lines are gathered together into a relatively small vicinity of the interface of two adjacent cells. Note that this occurs in the region of descending flows, where the flow has a hyperbolic point (in the region of ascending flows, on the contrary, the field decreases). Such a region of hyperbolicity represents a kind of attractor for the magnetic field. The magnetic field in this case grows exponentially, and its growth is limited by the finite conductivity of the medium. If the feedback effect of the field on the flow is neglected, stabilization occurs when the magnetic field reaches a value on the order of the square root of the magnetic Reynolds number $Re_m$ ~\cite{Kuznetsov20}. It is important to note that the growth of the magnetic field and its saturation in the filament at this stage can be determined by analyzing the behavior of the magnetic field at the free boundary (between convective zone and photosphere), which, according to observations, can be considered flat. Moreover, the formation of magnetic filaments does not depend on the internal structure of the convective cells, but is determined by the behavior of the convective flow along the free surface.

\subsection{Basic parameters of the solar convective zone}

Let us consider the main characteristics of the solar convection zone. According to observations, the horizontal size of convection cells, $L$, is approximately 500-1000 km. Data from \cite{soho}, as well as many others (see, for example,
the book \cite{eddy1979new}), yield values ??of about $1000 \mbox{ m/s}$ for the velocity $v$ in the cell, and a value of about $10^{-7} \mbox{ g/cm}^{3}$ for the density $\rho$ in the photosphere. Note also that there is no sharp gradient between the convection zone and the photosphere; the density changes smoothly (see, for example,
the review by \cite{kosovichev2009photospheric} and references therein). The density in the convective zone is, of course, greater than $10^{-7}\mbox{ g/cm}^{3}$. In the literature, the density at the boundary of the convective zone and the photosphere is considered to be on the order of $10^{-6}-10^{-5}\mbox{ g/cm}^{3}$. However, over the size of a convective cell, the density can be considered virtually constant and, accordingly, the flow in the cell itself is incompressible: $ \mbox{div}\mathbf{v}=0$. As for the average magnetic fields, their magnitude is on the order of several gauss (for our estimates, we will assume $B=10$ G). From this, we can estimate the ratio of the average convection kinetic energy density $<E_{kin}>$ and the magnetic field energy density $<E_m>=<B^2/(8\pi)>$. $<E_{kin}>$ exceeds the average magnetic field energy density $<E_m>$ by 2--3 orders of magnitude.

Based on these data, it follows that the behavior of magnetic fields during the formation of magnetic filaments in the solar convective zone can be described by a single equation — the induction equation, when the velocity field can be considered given (the so-called kinematic approximation):
\begin{equation} \label{MHD-1}
\frac{\partial {\bf B}}{\partial t}=\nabla \times ({\bf v} \times {\bf B})+\nu_m \Delta {\bf B};\,\,\,\mbox{div}\,{\bf v}=0.
\end{equation}
where $\nu_m=c^2/(4\pi\sigma)$ is the magnetic viscosity, $\sigma$ is the conductivity.

For the solar convective zone, the characteristic magnetic Reynolds number is of the order of $10^6$, which allows us to neglect the term with magnetic viscosity $\nu_m$ in (\ref{MHD-1}):
\begin{equation} \label{MHD-2}
\frac{\partial {\bf B}}{\partial t}=\nabla \times ({\bf v} \times {\bf B}).
\end{equation}
This equation is often called the frozenness equation. Frozenness (in this case, of the magnetic field) means that each fluid particle is pasted to its magnetic field line, i.e., the particle moves along with its magnetic field line. As is easily seen, each particle has only one degree of "freedom" — motion along a field line, in which, by virtue of the vector product in (\ref{MHD-2}), the magnetic field ${\bf B} $ remains constant. The velocity component ${\bf v_n} $, normal to the magnetic field vector, represents the velocity of the magnetic field line. Note that magnetic field lines can be compressed: in a general situation, $\mbox{div}\,{\bf v_n}\neq 0$. This is one of the main consequences of the frozenness equation (\ref{MHD-2}).

According to SOHO (Solar and Heliospherical Observatory) data, filaments in the solar convection zone are mostly concentrated near the boundary between convection cells and are oriented accordingly along the downward flows. SOHO observations also show the absence of filaments in the centers of \cite{soho} cells (see Fig. 1).

\begin{figure}
    \centering
    \includegraphics[width=10cm]{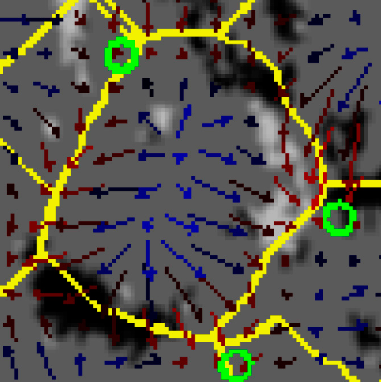}
    \caption{A solar cell (SOHO magnetogram). Red arrows indicate descending flows, blue arrows indicate ascending flows. Yellow lines indicate cell boundaries, and green lines indicate converging flows. Areas with the strongest magnetic fields are shown in black and dark gray. Courtesy of SOHO/MDI consortium. SOHO is a project of international cooperation between the ESA and NASA~\cite{soho})
    }.
    \label{fig:ionization}
\end{figure}

In this brief review, we will explain these experimental data. They are based on the frozen-in property of the magnetic field, as well as another key observation: the boundary of the convective zone—the free surface—can be considered flat with good accuracy. These two facts help clarify many experimental observations regarding filament dynamics during the first stage of filament formation. At the end of this stage, the magnetic field in the filament loses its frozen-in property due to finite conductivity. We will show that, in the kinematic approximation within the framework of equation (\ref{MHD-1}), the magnetic field in the filament saturates. Relative to the initial magnetic field, the maximum value of $B$ in the filament increases by a factor of $(Re_m)^{1/2}$. This increase is significant: for the solar convective zone, the magnetic field can increase by two to three orders of magnitude and reach values ??of $1-10$ kG.

With such a magnetic field enhancement, the filament's feedback effect on the flow can no longer be ignored. At this stage, with a sufficiently large magnetic field in the filament, the magnetic field pressure gradient begins to impede the convective flow, leading to a shift of hyperbolic regions toward the convective flow along the free surface. As a result, the transverse size of the filament increases and stops when the kinetic energy density and the magnetic field energy density become comparable. Since, within the kinematic approximation, the magnetic field enhancement is associated with the flow along the free boundary and, importantly, is not related to its internal structure, free-surface effects must also play a key role at this stage ~\cite{Kuznetsov24}.

In this paper, we show that the evolution of the magnetic field at the boundary of convective cells is a two-stage process. During the kinematic stage, the magnetic field increases due to the presence of a hyperbolic point in the flow, corresponding to the onset of the downward flow along the boundary between the cells. In the next stage, as the field increases, the magnetic pressure gradient becomes significant, leading to a shift of the hyperbolic point toward the convective flow along the free boundary. This leads to a decrease in the magnetic field in the filament compared to the value determined by the kinematic approximation. Note that a similar situation arises in problems of space gas dynamics related to dynamo theory ~ \cite{Chertkov99,Sokoloff15}, where the behavior of the magnetic field can also be studied for flows along the boundary of the convective zone.

The plan of this paper is as follows. In the second section, following mainly the work of \cite{Kuznetsov20}, we will consider the dynamics of magnetic filament formation in the kinematic approximation in two-dimensional ($x-y$ geometry) for a steady flow in the form of a periodic system of rollers. Analytical and numerical results will be presented. In particular, the role of hyperbolic flow regions (downdrafts) as a kind of attractor for the magnetic field will be demonstrated. The  next section examines the dynamics of magnetic filament formation in convective cells of arbitrary shape, based on an analysis of magnetic field behavior at the upper boundary of the convective zone within the kinematic approximation. This analysis is based on observational data that this boundary can be considered flat. In this case, at this stage, it is possible to elucidate the most general patterns of filament formation and development -- a process independent of the internal structure of convective cells ~\cite{Kuznetsov24}. The fourth section examines the feedback effect of emerging magnetic filaments on the convective flow. In this section, we restrict ourselves to a one-dimensional analysis of filament behavior at the boundary of the convective zone. Since the velocity in the hyperbolic region of the downflow is small (note that the velocity vanishes at the hyperbolic point), the magnetic pressure gradient stops the counterflow at a relatively early stage, leading to a shift in the hyperbolic point. As a result, the filament size increases compared to the filament width in the kinematic case. The final section is a conclusion.

\section{Evolution of the magnetic field in the kinematic approximation}

Let us first consider the generation of magnetic filaments in the simplest formulation within the framework of equation (\ref{MHD-1}) for a two-dimensional steady-state flow in the form of a periodic chain of convective rolls, assuming the initial magnetic field $\mathbf{B_{0}}$ to be uniform,
directed upward along the $y$ axis. The velocity will be represented through the stream function $\psi $: $v_{x}=-\partial
_{y}\psi ,\,\,v_{y}=\partial _{x}\psi $. For a periodic
chain of cells with a circulation that changes sign from one
cell to another, $\psi $ can be written as a product
of two sines: $\psi
=C\sin \left( k_{1}x\right) \sin \left( k_{2}y\right) $.
For a stationary system of cells for Benard convection, $k_{1}\neq k_{2}$, but
they are of the same order \cite{gershuni1977convective}
(see also \cite{kuznetsov1980weak}). For simplicity, we will set $k_{1}=k_{2}=1$, and the constant $C=1$. Then
\begin{equation}
\psi =\sin x\cdot \sin y,  \label{psi}
\end{equation}
and components of velocity
\[
v_{x} =-\sin x\cdot \cos y,\,\, v_{y} =\cos x\cdot \sin y.
\]
The line $y=0$ is considered the upper boundary
of convective zone. The velocity along the boundary in this case is parallel
to the surface, and the normal component is therefore zero.

In equation (\ref{MHD-2}) for the magnetic field in the ($x,y$) plane, we introduce the magnetic potential
\begin{equation}
A=-B_{0}x+a,  \label{fluct}
\end{equation}
dynamics of which can by found by integrating equation
\begin{equation}  \label{magnetic}
\frac{\partial A}{\partial t}+(\mathbf{v}\cdot \nabla) A=0,
\end{equation}
where the magnetic field is expressed as:
\begin{equation}  \label{field}
B_x= \frac{\partial A} {\partial y}, \,\, B_y=-\frac{\partial A} {\partial
x}.
\end{equation}
Hence, it follows, in particular, that the equipotential line $A=\mbox{const}$ coincides with the magnetic field line. Moreover, as follows from the equation (\ref{magnetic}), the motion of this line is determined only by the normal component of the velocity $\mathbf{v}$, which is a direct consequence of the frozenness  of the magnetic field. From this, it is easy to understand the dynamics of the magnetic field lines (see Fig. 2).
\begin{figure}
    \centering
    \includegraphics[width=10cm]{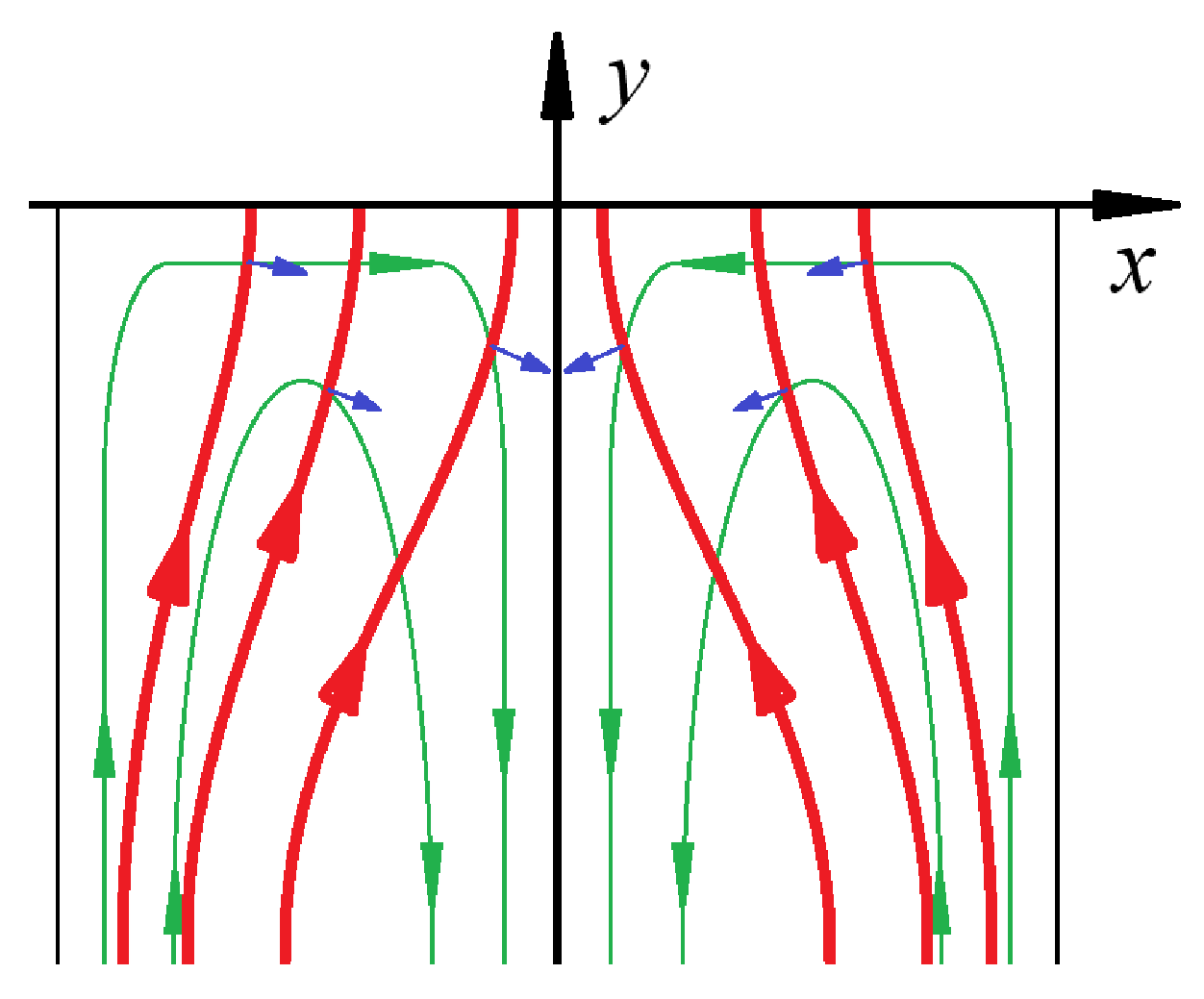}
    \caption{Magnetic field lines (red) in a convective cell (current lines are marked in green), black arrows (perpendicular to the magnetic field lines) show the direction of movement of the lines}
\end{figure}

In the numerical experiment, a rectangular
region [$-\pi \leq x\leq \pi ,\,0\geq y\geq
-\pi $] was chosen. Along the lines $x=\pm \pi ,\,0\geq y\geq -\pi $, the fluid
rises (upward flow), while along the central line
$x=0,\,0\geq y\geq -\pi $, it sinks (downward flow). If, at
the initial instant of time, the magnetic field $\mathbf{B_{0}}$ is directed along
the $y$ axis, then the normal component
of the velocity with respect to $\mathbf{B_{0}}$ will have positive
$x$-projections in the upper right corner in the left cell, i.e. to the line $x=0,\,0\leq y\leq \pi /2$ and, correspondingly, the negative
$x$-projections in the left corner of the right cell. As will be shown later, the magnetic
field lines in this region will be attracted to the downward
flow, i.e., to the central line where the filament is formed.

The equation for $A$ is easily integrated
using the method of characteristics. The equations for the characteristic are
\[
\frac{d\mathbf{r}}{dt}=\mathbf{v}(\mathbf{r})
\]
with initial conditions $\mathbf{r}|_{t=0}=\mathbf{\alpha}$. These equations
when written component-wise represent the Hamilton equations
\begin{equation}
\frac{dx}{dt}=-\frac{\partial \psi }{\partial
y},\,\,\frac{dy}{dt}=\frac{\partial \psi }{\partial x}  \label{Hamilton}
\end{equation}
with initial conditions $x(t=0)=\alpha_{x}$ and $y(t=0)=\alpha_{y}$.
The coordinates $x$ and $y$ in these equations are canonically
conjugate quantities, and the stream function $\psi (x,y)$ is a
Hamiltonian. Since the velocity field is independent of
time, $\psi (x,y)=\mbox{const}$. Thus, the dynamics of the system (\ref{Hamilton})
is determined by the properties of the Hamilton function $\psi (x,y)$. For a bounded domain,
the function $\psi (x,y)$, like a two-dimensional relief, is characterized by its extrema—
minima, maxima, and saddle points. At the extremum points,
the gradient of $\psi (x,y)$ is zero, which corresponds to zero velocity. The question of what the extremum point is
is determined from the expansion of $\psi (x,y)$ in the neighborhood of the extremum ${\bf r}={\bf r_0}$:
\begin{equation}
\psi({\bf r})= \psi({\bf r_0}) +\frac 12 D_{ij}\Delta x_i \Delta x_j + ...,\label{extremum}
\end{equation}
where $\Delta {\bf r}= {\bf r}-{\bf r_0}$,
\[
D_{ij}=\frac{\partial^2\psi }{\partial x_i \partial x_j}|_{\bf r=r_0}.
\]
At its maximum or minimum, the quadratic form $D_{ij}\Delta x_i \Delta x_j$ is sign-definite. At this point,
the eigenvalues ??of the matrix $D_{ij}$ are sign-definite if the following inequality holds:
\begin{equation}
\label{Okubo}
\psi_{xx}\psi_{yy}-\psi^2_{xy}>0.
\end{equation}
According to \cite{weiss1991dynamics,okubo1970horizontal}, such points are called elliptical, and the regions where inequality (\ref{Okubo}) holds are called elliptical. With the opposite sign in inequality (\ref{Okubo}), the stationary point becomes hyperbolic, and the region with the opposite sign in (\ref{Okubo}) is called hyperbolic.

For the stream function (\ref{psi}), the point $x=y=0$ is hyperbolic; according to Figure 2, the magnetic field should be attracted to it, leading to magnetic field filamentation.

Now let's consider the solution to the problem for the initial condition (\ref{fluct}) for the stream function. Since the magnetic potential fluctuations are $a=0$ at $t=0$, the magnetic potential on the characteristic
will be $A=-B_{0}\alpha_{x}$, i.e., it depends only on the initial
value of the $x$-coordinate of the fluid particles $\alpha_{x}$.

The equations (\ref{Hamilton}) are easily integrated. From
the equation $\psi(x,y)=\psi(\alpha_x,\alpha_y)$, one can find,
for example, $y=y(x, \alpha_x,\alpha_y)$ and then
substitute this dependence into the right-hand side of the first
equation (\ref{Hamilton}). The resulting equation for $x$
\[
\frac{d x}{dt}=v_{x}(x,a_{x},a_{y})
\]
integrates trivially. Thus, we come to a common solution
Cauchy problem for the equation (\ref{magnetic}), which is written
in an implicit form.

We will be interested in the behavior of the maximum magnetic field.
The qualitative considerations given above show that the maximum
magnetic field must be in a small vicinity
points $x=y=0$, i.e. for the center of the beginning of the downward flow, at the boundary
between cells. Let's consider small deviations from the point $x=0$, $y=0$,
considering $x$ and $y$ small. For such values of $x$ and $y$, the stream function in
according to (\ref{psi}) can be approximately written as
$\psi ={x}{y}$ with initial condition
$\psi ={\alpha_{x}}{\alpha_{y}}$.
For such a stream function, the equation
for ${x}$ becomes linear:
\[
\frac{dx}{dt}=-{x},
\]
the solution of which yields an exponential
narrowing of scale
\begin{equation}
x=\alpha_{x}e^{-t}.  \label{compression}
\end{equation}
$y$ exhibits exponential
growth: $y=\alpha_{y}e^{t}$. These asymptotics determine the behavior
of the magnetic field in this region.
For ${x\rightarrow 0}$ and ${y\rightarrow 0}$, it is easy to see that
\begin{equation}
B_{x}=0,\,\,B_{y}=B_{0}e^{t}.  \label{filam}
\end{equation}
If $B_{x}$ is not zero at the initial time, it decays exponentially with time. Thus, the maximum magnetic field value grows exponentially with time in the vicinity of the hyperbolic point. Importantly, the maximum field is directed along the downdraft. In this case, the horizontal component of the magnetic field tends to zero. As will be shown in the next section, this situation is typical: the magnetic field in the filaments at the boundary between two convective cells is normal to the free surface and can be directed either vertically upward or vertically downward, as shown in Fig. 1 (dark and light regions).

We now estimate the contribution of finite conductivity to the field evolution. In dimensionless form, equation (\ref{MHD-1}) for the field ${\bf B}$ has the form
\begin{equation} \label{Re}
\frac{\partial {\bf B}}{\partial t}=\nabla \times ({\bf v} \times {\bf B})+\frac{1}{\mbox{Re}_{m}} \Delta {\bf B}.
\end{equation}
At $t\to \infty$, due to magnetic viscosity, the magnetic field is saturated. Obviously, in order to find a stationary field, it is sufficient to consider the stationary equation (\ref{Re}) for $B_{y}$ for $y=0$:
$$\frac{\partial}{\partial x}  \left(B_{y} \sin  x+ \frac{1}{\mbox{Re}_{m}} \frac{\partial B_{y}}{\partial x} \right)=0.$$
Integrating this equation gives:
$$B(x)=C \exp \left( \mbox{Re}_{m} \cdot \cos x \right).$$
This solution describes a field that has a maximum near zero. For large values ??of $\mbox{Re}_{m}$, it is quite narrow, and the field is described by the expression:
$$B(x)=C \exp \left[\mbox{Re}_{m} \left( 1-\frac{x^{2}}{2}\right)\right].$$
The constant $C$ is found from the condition of conservation of the magnetic flux through the free surface, from which the maximum value of the magnetic field is determined by the expression: $B =B_{0}\left(2 \mbox{Re}_{m}\pi ??\right)^{1/2}.$ Thus, we have an increase in the magnetic field by $\left(2 \mbox{Re}_{m}\pi\right)^{1/2}$ times. The width of the maximum will be inversely proportional to $\sqrt{Re_m}$.

Numerically, the magnetic field ${\bf B}$ for the velocity field given by the stream function (\ref{psi}), and ${\bf B}(t=0)={\bf B}_0$ was found by solving the equation for $a(x,y,t),$ with the boundary conditions
$$a|_{x=-\pi}=a|_{x=\pi};\,\,{\partial a}/{\partial y}|_{y=0}=a|_{y=-\pi}=0.$$
The result of the numerical solution is shown in Fig. 3 for $\mbox{Re}_m=500$.
\begin{figure}
    \centering
    \includegraphics[width=10cm]{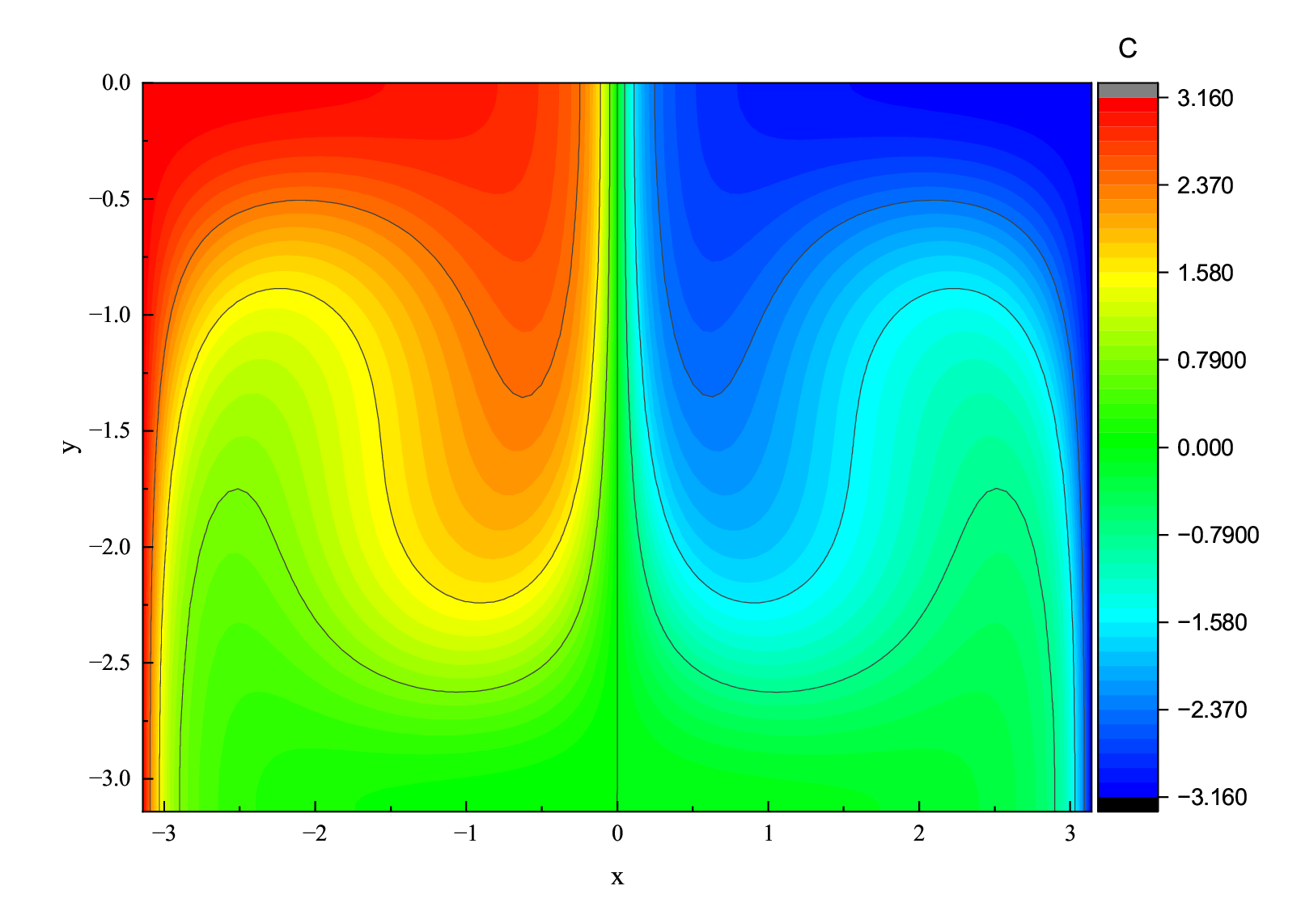}
    \caption{The magnetic field potential at $t=3$ is $\mbox{Re}_m=500$.
    }
\end{figure}
The dependence of the field on time at different magnetic Reynolds numbers is shown in Fig. 4.

\begin{figure}
    \centering
    \includegraphics[width=10cm]{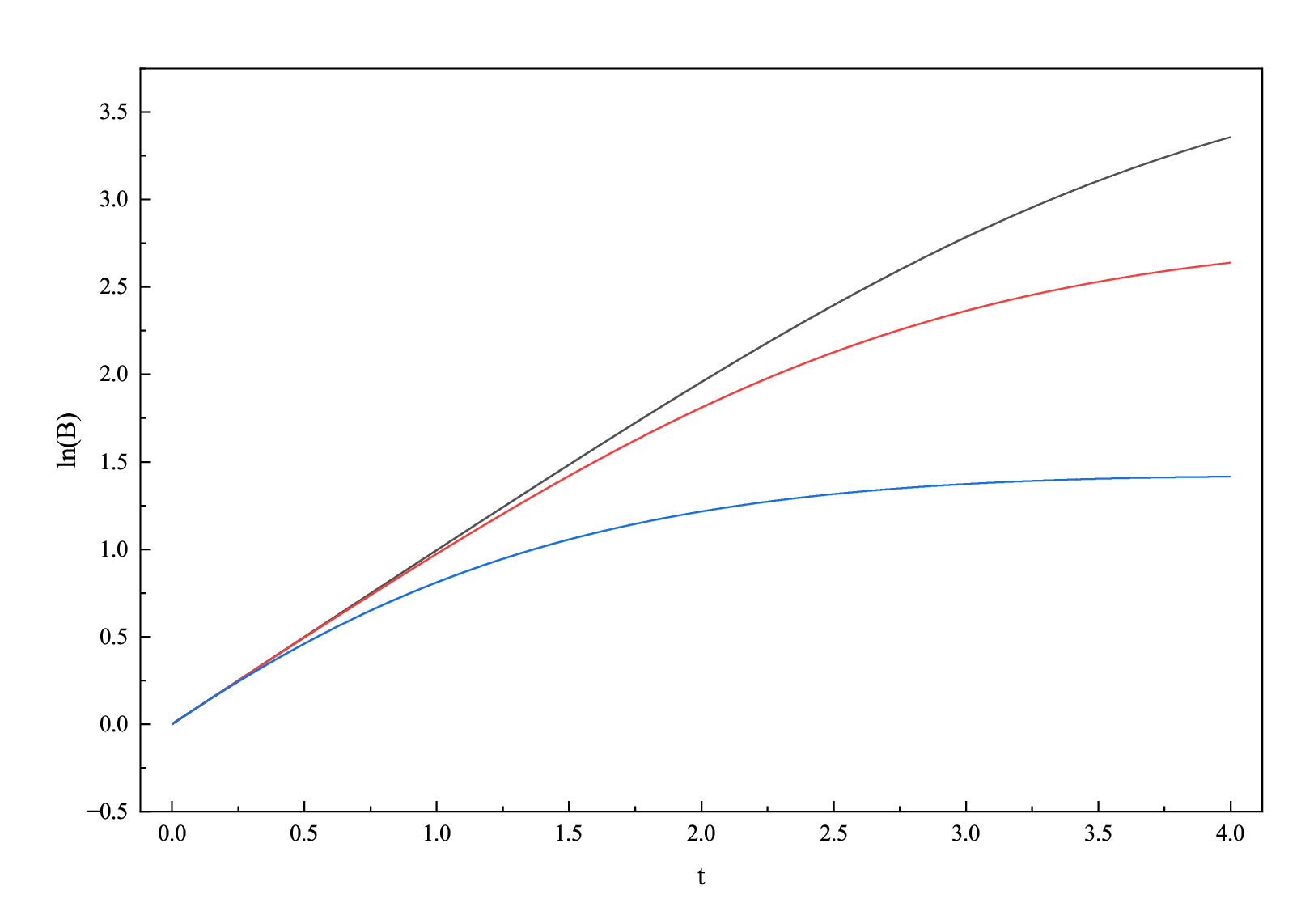}
    \caption{Dependence of the maximum magnetic field at $x=y=0$ on time. The blue line corresponds to $\mbox{Re}_{m}=5,$ the red line to $\mbox{Re}_{m}=50,$ the black line to $\mbox{Re}_{m}=500.$
    }
\end{figure}
It should be noted that in the absence of dissipation, the maximum value of the magnetic field grows exponentially with an increment of $\gamma=1$. This value of $\gamma$ coincides with $\partial_xv_x$ at $x=y=0$, where $v_x=\sin x$.

\section{Formation of magnetic filaments in convective cells of arbitrary shape}

This section examines the dynamics of magnetic filament formation in convective cells of arbitrary shape, based on an analysis of magnetic field behavior at the upper boundary of the convective zone within the kinematic approximation. This analysis is based on observational data indicating that this boundary can be considered flat. We will first assume that the magnetic field dynamics are determined by the frozen-in equation (\ref {MHD-2}). The results presented in the previous section demonstrate that, in the two-dimensional case, magnetic filaments emerge and develop at the free boundary of the convective zone in the vicinity of hyperbolic points corresponding to downward flows. As we will see in this section, this tendency holds for arbitrary convective cell structures, regardless of their internal structure, and is determined by the behavior of the magnetic field at the boundary of the convective zone. Since the boundary can be considered flat, the normal component of the velocity at the boundary, $v_n$, should be set equal to zero. Then the equation for the normal component of the magnetic field at the boundary $\Gamma$ within the framework of equation (\ref{MHD-2}) turns out to be autonomous:
\begin{equation}
\label{surf}
\frac{\partial B_n}{\partial t}+ ({\bf v_{\perp}}\cdot\nabla_{\perp})B_n= -B_n (\nabla_{\perp}\cdot{\bf v_{\perp}}).
\end{equation}
In particular, for two-dimensional flows this equation is written as
\begin{equation}
\label{surf-1}
\frac{\partial {B}_y}{\partial t}+ v_{x}\frac{\partial B_y}{\partial x}= -B_y\frac{\partial v_x}{\partial x}.
\end{equation}
Thus, the dynamics of the normal component of the magnetic field on the free surface is determined only by the tangential component of the velocity on $\Gamma$ and does not depend on the convection structure in the inner convection region.

Both equations (\ref{surf}) and (\ref{surf-1}) admit solutions by the method of characteristics along the characteristic $dx/dt=v_x$; the component $B_y$ obeys the equation
\[
\frac{dB_y}{dt}= - B_y \frac{\partial v_x}{\partial x}.
\]
Thus, $B_y$ is transported by advection with velocity $v_x$, simultaneously decaying or growing depending on the sign of $-\partial v_x/\partial x$. At the center of the convective cell, the value of
$-\partial v_x/\partial x$ will be negative (note that at this point $v_x=0$), and $B_y$ will decrease (exponentially in time for stationary flows). In the region of downdrafts, the normal component of the magnetic field $B_y$ will grow exponentially in time for a stationary convective flow with an growth rate of $\gamma=\partial_xv_x$, which is completely consistent with the previous analysis ($v_x=\mbox{sin}x$).

As for the horizontal component $B_x$, it is easy to understand that it will decay over time, exponentially for a stationary flow. This follows, in particular, from the equation for $B_x$ along the downward flow at the boundary between convective cells, since the equation for this component, up to a change in notation, will coincide with
(\ref{surf-1}). Since at the beginning of this flow, the velocity is directed downward and has a negative derivative of the velocity with respect to $-y$, we find that $B_x$ will decrease in magnitude. From this consideration, it follows that the formation of magnetic filaments is a fairly general (rough) phenomenon, independent of the flow structure in the inner region of the convective flow. Importantly, at the boundary, the magnetic field in the filament is directed vertically. This same conclusion, as we will see below, is also valid for the behavior of the magnetic field at the two-dimensional boundary $\Gamma$, i.e., for (\ref{surf}).

Like equation (\ref{surf-1}), (\ref{surf}) describes the advection of $B_n$ by a horizontal flow ${\bf v_{\perp}}$, where the equations for the characteristic
\begin{equation}
\label{char}
\frac{d {\bf r}}{dt}={\bf v_{\perp}}.
\end{equation}
From the center ${\bf r}={\bf r}_0$, where ${\bf v_{\perp}}=0$, the flux along the surface will spread out until it reaches the boundary of the convective cell, where ${\bf v_{\perp}}=0$ (this entire boundary is a whole hyperbolic line). It is easy to understand that
at the center $\mbox{div}\,{\bf v_{\perp}}$ is positive (source), and at the boundary of the convective cell $\mbox{div}\,{\bf v_{\perp}}$ is negative
(sink). Thus, the normal component of the magnetic field along the characteristic -- the solution of equation (\ref{char}) -- will change in accordance with:
\[
\frac{dB_n}{dt}= - B_n \mbox{div}\,{\bf v_{\perp}}.
\]
It follows that the normal component of the magnetic field $B_n$ will decay as it moves away from the center and increase correspondingly as it approaches the boundary of the convective cell, where magnetic filaments are formed. The greatest growth and, accordingly, the largest filament should occur along the boundary, where the value of $-\mbox{div}\,{\bf v_{\perp}}$ is maximum. Just as for two-dimensional flows described by equation (\ref{surf-1}), based on the same considerations about the attenuation of the $B_x$ component, we can conclude that the horizontal component of the magnetic field on the free surface, at least when approaching the boundary of the convective cell, should decay, which suggests that in the filaments that arise at the boundary of the convective cell, the magnetic field should be normal to the free surface. If the flow were stationary, the growth of filaments would be exponential in time. Along the boundary itself,
the filament with the maximum value of $-\mbox{div}\,{\bf v_{\perp}}$ should exhibit the greatest growth. Other filaments should form near the maximum.

Based on this entire consideration, it can be concluded that, in a general situation, filaments should concentrate at the boundaries between cells, which is confirmed by observations (see, for example, Fig. 1). Furthermore, as observations indicate, magnetic filaments are quite rare in the cell centers: due to their frozenness state, they should be carried toward the boundaries of convective cells. Regarding magnetic field saturation due to finite conductivity, if the kinematic approximation criterion is met, the magnetic field growth is of the order of $\sqrt{Re_m}$. At magnetic Reynolds numbers of the order of $10^6$, the estimate for the magnitude of the magnetic field in the filament can be a large value of $1 - 10$ kG.

\section{The feedback of the magnetic field on the flow}

Before considering the feedback of growing magnetic filaments, we will make a number of important remarks that will allow us to study this issue. Firstly, at the kinematic stage, the generation of filaments occurs in a small vicinity of the hyperbolic points corresponding to the downward flows. At these points the speed of the convective flow becomes zero. It is for this reason that at the kinematic stage these regions represent attractors for magnetic fields. The width of the filaments in this regime at high magnetic Reynolds numbers is small, on the order of $Re_m^{-1/2}$. Secondly, we have shown that at this stage, to study the formation of filaments in a general situation, it is enough to consider the behavior of magnetic fields on the free surface - the boundary of the convective zone. Moreover, we found out that the horizontal component of the magnetic field in this case is small; for two-dimensional stationary flows, this component decays exponentially in the region of filament localization. In other words, at the next stage, when it is necessary to take into account the feedback of the growing filament on the convective flow, it is natural to neglect this component.

Both of these circumstances play an important role in the study of the feedback of magnetic filaments on convection.

Let us write the Navier-Stokes equation in the Boussinesq approximation for the velocity ${\bf v}$ taking into account pondermotive forces:
$$\frac{\partial {\bf v}}{\partial t}+\left( \textbf{v}\cdot \nabla \right)\textbf{v}=$$
\begin{eqnarray} \label{NSBus}
= -\frac{1}{\rho} \nabla p+\nu Ra T \textbf{e}_{y}-\frac{1}{4 \pi \rho} \left( \textbf{B} \times \left( (\nabla \times \textbf{B} \right) \right)+\nu \Delta \textbf{v}.
\end{eqnarray}
Here $\rho$ is the density, $Ra$ is the Rayleigh number, $p$ is the pressure fluctuation, $T$ is the difference between the local temperature and the average temperature, linearly dependent on $z$, and $\nu$ is the kinematic viscosity. The velocity is assumed to be locally incompressible:
$\mbox{div}\,{\bf v}=0$. From now on, the magnetic field will be measured in Alfven units: $B/\sqrt{4\pi\rho}\to B$. Throughout this section, unlike the notation in Section 2, the $z$ axis is directed vertically upward.

For equation (\ref{NSBus}), boundary conditions must be set. On the free boundary $z=0$, they are represented by the equalities
\begin{equation}
\{\sigma_{ik}n_k\}=0,
\label{BC-G}
\end{equation}
where $\sigma_{ik}$ is the stress tensor, the curly brackets $\{ ...\}$ denote a jump, and ${\bf n}$ is the unit normal vector to the boundary (${\bf n}^2=1$). At $z=0$, $T$ is assumed to be zero.

Further, we restrict our consideration to two-dimensional flows, for which the boundary conditions (\ref{BC-G}) are written as
\begin{equation}
\{\sigma_{xz}\}=0,\,\,\{\sigma_{zz}\}=0,
\label{BC-2D}
\end{equation}
where
$$\sigma_{xz}=-v_{x}v_{z}+B_{x}B_{z}+\nu\left(\frac{\partial v_{x}}{\partial z}+\frac{\partial v_z}{\partial x}\right);$$
$$\sigma_{zz}=-v_{z}^{2}-\left(p+\frac{B^2}{2}\right)+B_{z}^{2}+2\nu\frac{\partial v_{z}}{\partial z},$$
and $z$ means vertical coordinate.

As noted above, we assume the vertical component of the velocity $v_{z}$ on the surface to be zero. We also neglect the horizontal component of the magnetic field at $z=0$ due to its smallness compared to the normal $B_z$, which is valid for $\mbox{Re}_{m}\gg 1$. At the boundary, $B_z$ is assumed to be continuous. As a result, the boundary conditions (\ref{BC-2D}) take the form:
$$\frac{\partial v_{x}}{\partial z}|_{z=0}=0;$$
$$\left(-p+2\nu\frac{\partial v_{z}}{\partial z} \right)|_{z=0}=0.$$
Substituting these relations into the equation (\ref{NSBus}),
for the horizontal component of the velocity ($v_{x}$) at $z=0$ we obtain the equation
\begin{equation} \label{dyn}
\frac{\partial v_{x}}{\partial t}+v_{x}\frac{\partial v_{x}}{\partial x}=-B_{z}\frac{\partial B_{z}}{\partial x}+3 \nu \frac{\partial^{2} v_{x}}{\partial x^{2}}+\nu \frac{\partial^{2} v_{x}}{\partial z^{2}}.
\end{equation}
In this equation, formally, only the second term on the right hand side describes the influence of the volumetric part of convection on the structure of the flow on the surface $z=0$. The most important term in this equation is the first term on the right hand side, which represents the pressure gradient of the magnetic field: $-\partial_x B_z^2/2$. This is due to the fact that magnetic filaments at the kinematic stage are generated at a hyperbolic point where the velocity is zero; it is small in the region of filament localization at $\mbox{Re}_{m}\gg 1$. If this is true, then it is clear that the magnetic field pressure gradient has the greatest influence on convective motion. In order to find out its role instead of the equation (\ref{dyn}), we carried out our hypothesis for the equation
\begin{equation}
\label{dyn-1}
\frac{\partial v}{\partial t}+v\frac{\partial v}{\partial x}=-B\frac{\partial B}{\partial x}+3Re^{-1} \frac{\partial^{2} v}{\partial x^{2}}.
\end{equation}
In this equation, $v$ is the velocity along the surface ($\equiv v_x$), and the magnetic field $B$ ($\equiv B_z$) is normalized to the characteristic value of the velocity $U$. Compared to (\ref{dyn}), we have neglected the last term, which, in our opinion, is not significant at the initial stage of magnetic filament interaction, since the characteristic vertical size is comparable to the size of the convective cell. As we will see, everything develops on much smaller scales compared to the size of the convective cell.

Equation (\ref{dyn-1}) is completed by the equation for the normal component of the magnetic field $B$ ($\equiv B_z$).
\begin{equation} \label{eq:3-1}
\frac{\partial B}{\partial t} +  v\frac{\partial B}{\partial x}= -B \frac{\partial v}{\partial x} +\mbox{Re}_{m}^{-1} \frac{\partial^{2}B}{\partial x^2}\end{equation}

To ensure that the velocity $v$ changes slowly compared to the magnetic field $B$, in the numerical experiments conducted in this paper, the ratio between Reynolds numbers  was $10^2$: $Re=30$, and the magnetic Reynolds number $\mbox{Re}_{m}=1000$, which allowed us to identify the main patterns of magnetic filament growth. According to \cite{Sokoloff, magnet}, such a difference in Reynolds numbers is typical for the solar convective zone.

The numerical modeling was performed for equations
(\ref{dyn-1}) and (\ref{eq:3-1}), written in dimensionless variables. As the numerical modeling shows, the main effect of the feedback influence of the growing magnetic filament will be associated with a small shift of the hyperbolic point in the direction opposite to the flow along the boundary. The boundary conditions for the velocity and magnetic field were assumed to be cyclic, with $B|_{x=-\pi}=B|_{x=\pi},$ $v|_{x=-\pi}=v|_{x=\pi}.$

Integration of the system (\ref{dyn-1})--(\ref{eq:3-1}), over the interval $-\pi \le x \le \pi$, was performed numerically using an implicit difference scheme implemented by the sweep method, a special case of the Gaussian method for a tridiagonal matrix. The grid step along the $x$ coordinate was $h=\pi\cdot 10^{-5},$ along time $\Delta t=10^{-5}$. The following approximate initial conditions were chosen, which are essentially the first Fourier harmonic for the velocity, periodic in space:
\begin{equation} \label{IC}
v(x,0)=-\mbox{sin}\,x, \,\, B(x,0)\equiv B_0=0.2.
\end{equation}
In this case, the point $x=0$ separates two adjacent convective cells, and at the same time (for a two-dimensional flow) this point is a hyperbolic point. For the given initial conditions, the ratio of the average densities of the kinetic energy $\rho <v^2>/2$ and the magnetic field energy $B_0^2/(8\pi)$ was $12.5$. It follows that, at the initial stage, the dynamics of the magnetic field can be described within the kinematic approximation, i.e., using equation (\ref{eq:3-1}) at a given velocity (\ref{IC}). In this case, it immediately follows from this equation that the magnetic field will increase, reaching its maximum value at the point $x=0$. In the kinematic approximation, this growth will be exponential with an increment of $\gamma=1$. Within this approximation, due to finite magnetic viscosity, the growth will stop, reaching its maximum amplitude in the steady state (see \cite{Kuznetsov20, Kuznetsov24}).
\begin{figure}
    \centering
    \includegraphics[width=10cm]{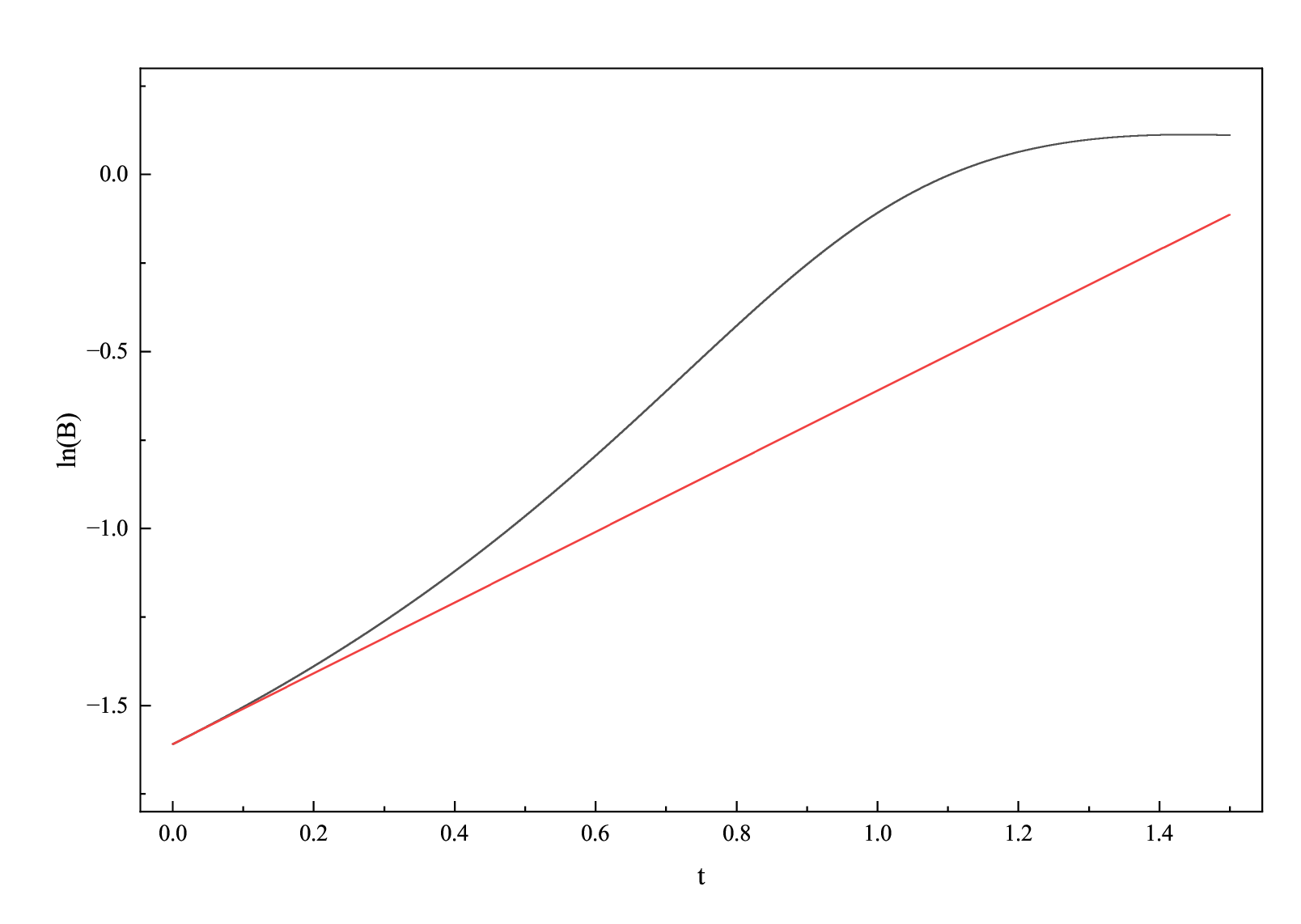}
\caption{Field dependence on time. The red curve shows $B_{max}$ in the absence of feedback, the black one shows the case of $Re=30$, $Re_{m}=1000$.
    }
\end{figure}

Figure 5 shows the dependences of the maximum magnetic field in the filament (a logarithmic scale is used).
\begin{figure}
    \centering
    \includegraphics[width=10cm]{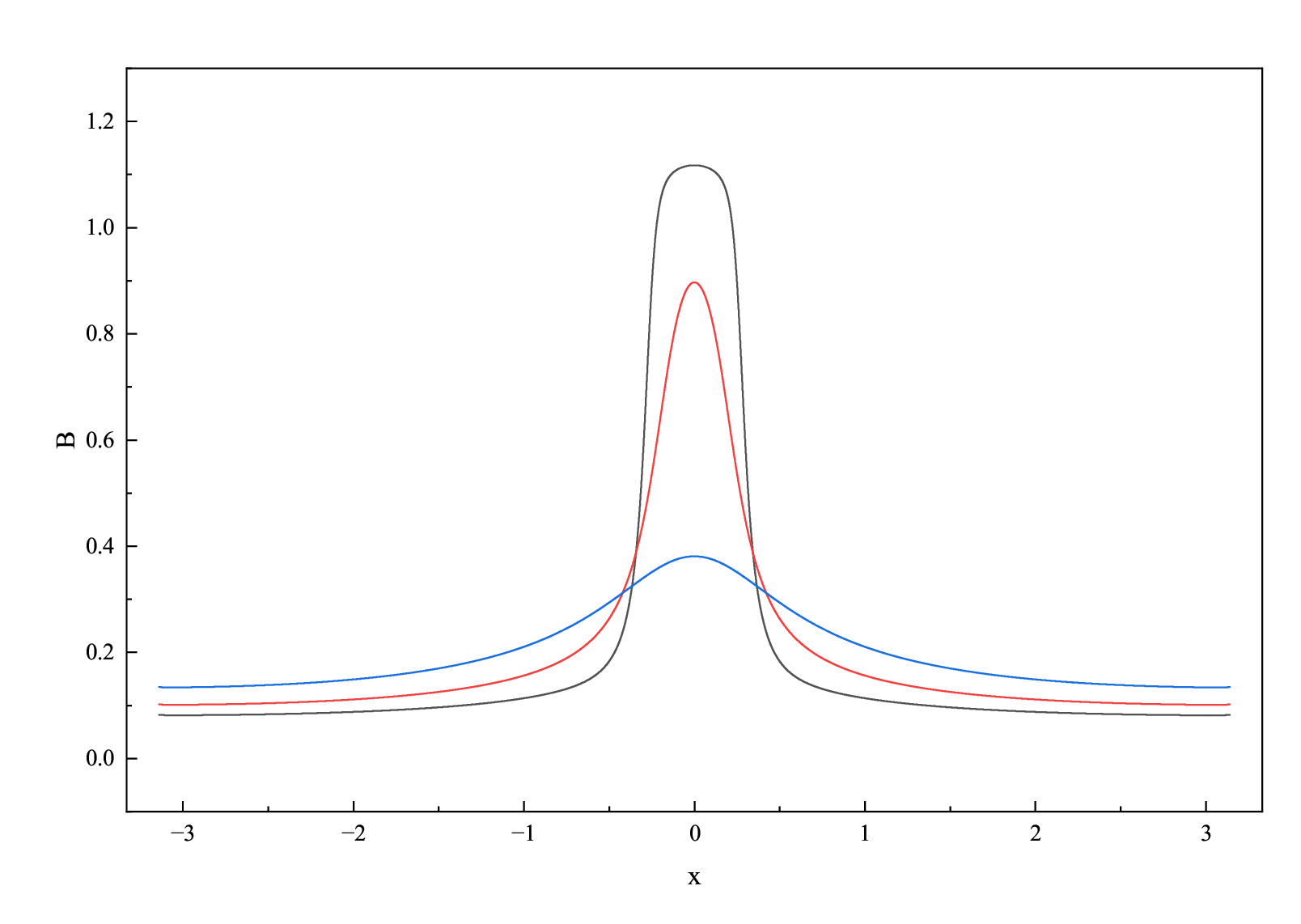}
    \caption{Magnetic field dependence on coordinate at $\mbox{Re}_{m} = 1000$, $Re = 30$. The black curve shows the dependence at $t = 0.5$, the red one $t = 1$, the blue one $t = 1.5$.
    }
\end{figure}
The red curve corresponds to the exponential growth of $B_{max}$ with $\gamma=1$ (kinematic approximation); the black curve corresponds to $Re_m=1000$, at short times when saturation is not yet observed).

\begin{figure}
    \centering
    \includegraphics[width=10cm]{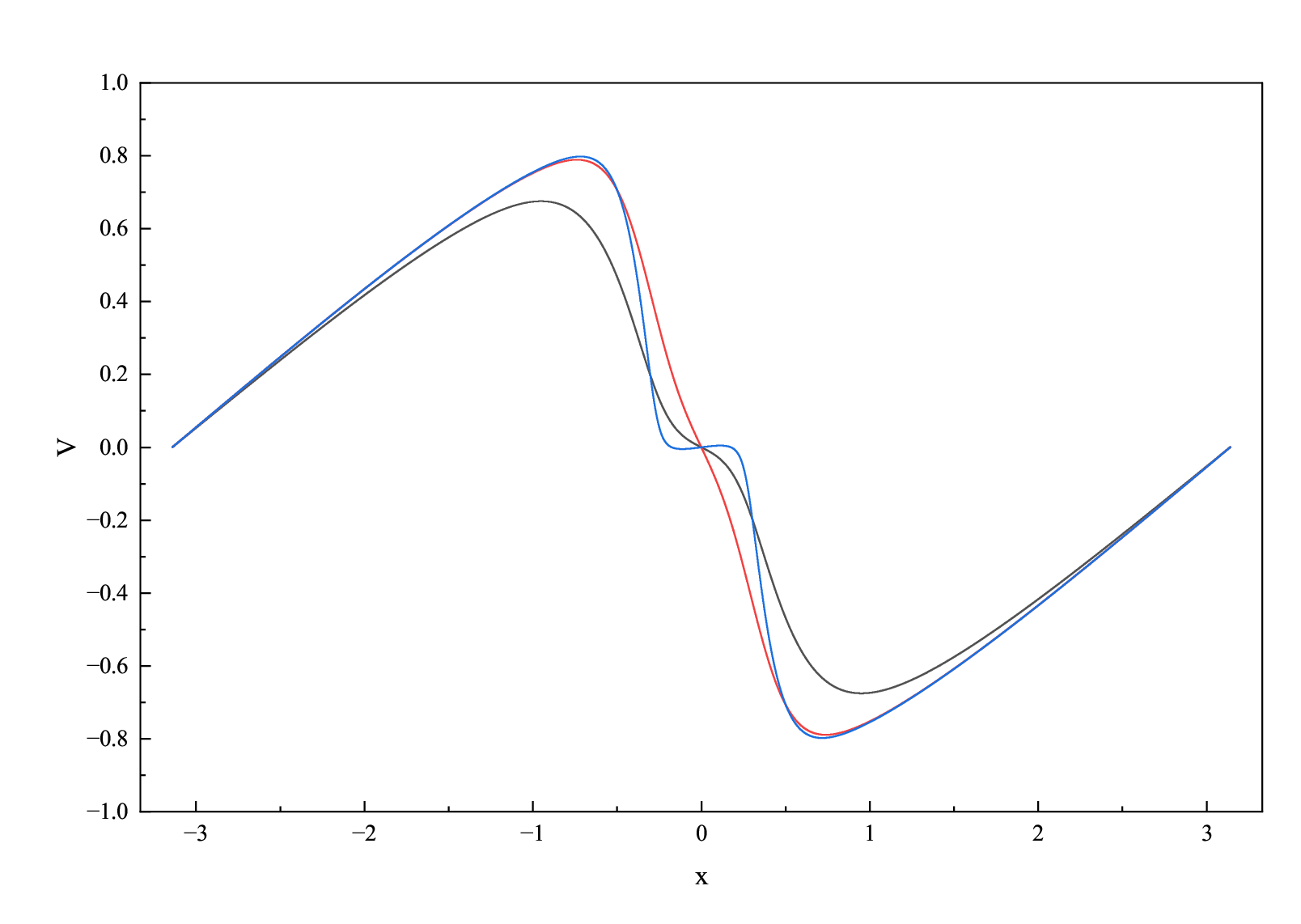}
    \caption{Dependence of velocity on coordinate at ${Re}_{m} = 1000$, $Re = 30$. The black curve shows the dependence at $t = 0.5$, the red one $t = 1$, the blue one $t = 1.5$.
    }
\end{figure}

The result of numerical integration of the system (\ref{eq:3-1},\ref{dyn-1}) for $Re=30$ and $Re_m=1000$ (black curve) is shown. It is evident that the local increment in this case as a function of time is greater than $\gamma=1$. Moreover, the deviation from the solid red curve initially increases with time, but then, at a time of about $t=1$, everything changes. The reason for this behavior is related to the influence of magnetic pressure, which displaces the flow from the filament region. In the kinematic case, we recall, the exponential growth is determined by the velocity gradient (in this case, at $x=0$).

As the filament forms, the magnetic pressure impedes the convective flow, pushing it to the filament's periphery from this point. In other words, the hyperbolic point shifts toward the flow, forming a plateau with zero velocity in the velocity profile. At the same time, the velocity gradient $\partial v/\partial x$ at the boundary increases, exceeding a value of $-1$. As a result, the local increment is greater than $\gamma=1$, which explains the behavior of the red curve in Figure 5.

Figure 6 shows the dependence of the magnetic field on the coordinate. Hence it can be seen that the filament has sharper boundaries, which correlate with the velocity dependences at different times, presented in Fig. 7.

\section{Conclusion}

Thus, this review shows that:

\begin{enumerate}

\item Convective cells are responsible for the formation of magnetic filaments in the solar convection zone;

\item At the kinematic stage, when the average kinetic energy exceeds the average magnetic field energy, the formation of magnetic filaments is due to the compressibility of continuously distributed magnetic field lines, a consequence of the frozen-in magnetic field. This process is exponential in time at high magnetic Reynolds numbers;

\item In the kinematic limit, the magnetic field saturation of the filaments occurs due to finite conductivity. The increase in the magnetic field in a filament is of the order of $\sqrt{Re_m}$. For the solar convection zone, the magnetic field values in filaments in this regime can be $1-10$ kG;

\item
Magnetic filaments on the Sun form at the boundaries of convective cells, which coincide with descending convective flows. The region of these flows represents a unique attractor of magnetic fields.
This attractor is a hyperbolic region of hydrodynamic flows. The magnetic field reaches its maximum at the hyperbolic point;

\item Based on observations, the boundary of the convective zone can be considered flat with good accuracy. In this case, the kinematic stage allows us to elucidate the most general patterns of filament formation and development — a process independent of the internal structure of the convective cell flow. Specifically, it has been shown that all types of local magnetic structures drift toward the interfaces of convective cells, leading to the formation of magnetic filaments. These theoretical results explain a number of observational data from the SOHO mission (Fig. ~\ref{fig:ionization});

\item
When the magnetic field growth in the filaments is significant, the growing magnetic fields in the filaments begin to influence the convective flow. At this stage, we found that the magnetic pressure gradient pushes the convective flow, shifting the hyperbolic point toward the flow. As a result, the filament size increases compared to the filament width in the kinematic case. This process is estimated to stop when the kinetic energy density and the magnetic field energy density become comparable.

\end{enumerate}

\end{document}